# Verification of Distributed Artificial Intelligence Systems in Bioinformatics

Aedin Pereira, Julia Ding, Zaina Ali, Rodion Podorozhny

## 1  Abstract

Software is a great enabler for a number of projects that otherwise would be impossible to perform. Such projects include Space Exploration, Weather Modeling, Genome Projects, and many others. It is critical that software aiding these projects does what it is expected to do. In the terminology of software engineering, software that corresponds to requirements, that is does what it is expected to do is called correct. Checking the correctness of software has been the focus of a great deal of research in the area of software engineering. Practitioners in the field in which software is applied quite often do not assign much value to checking this correctness. Yet, as software systems become larger, potentially combined with distributed subsystems written by different authors, such verification becomes even more important. Concurrent, distributed systems are prone to dangerous errors due to different speeds of execution of their components such as deadlocks, race conditions, or violation of project-specific properties. This project describes an application of a static analysis method called model checking to verification of a distributed system for the Bioinformatics process. In it, we evaluate the efficiency of the model checking approach to the verification of combined processes with an increasing number of concurrently executed steps. We show that our experimental results correspond to analytically derived expectations. We also highlight the importance of static analysis to combined processes in the Bioinformatics field.

## 2  Introduction

A multi-agent system is a type of Distributed Artificial Intelligence(DAI) system containing several agents that draw information from their individual environments and communicate it with each other to evaluate the best way to collaboratively complete a task. Examples of multiple agents in such systems include communicative agents implemented through negotiation in finite-state machines. These communicative agents exchange data, while non-communicative agents observe behavior without transmitting data to other agents. A DAI system consists of properties that represent transitions from states and their propositions to succeeding states and their respective propositions.

In many scientific fields, especially in Bioinformatics, precision and accuracy are imperative. Even the smallest mistakes can cause disastrous consequences with large-scale impacts. Thus, a need for verifying the correctness of systems and processes is created. While the verification of the integrity of an entire system as a whole is too large for practical purposes, one can largely verify a system's effectiveness and correctness by checking certain key properties against how the system operates. Conducting verification on such a set of properties can potentially isolate and eliminate virtually all of the flaws in the system.

The objectives of this project are to map the genome annotation process to a Kripke structure then, introduce errors into the Kripke structure and utilize a Computational Tree Logic (CTL) model checker to flag these errors. Moreover, we can perform an analysis of the run time of the CTL model checker and draw conclusions about the computational complexity of CTL model checking which is an important factor in gauging if it is a viable method to validate the use of DAI in the genome process.

## 3    Background Information

### 3.1    Overview of CTL model checking

One approach to such verification is model checking, in which the system is modeled through a Kripke structure, an example of which is shown below in Figure 1.

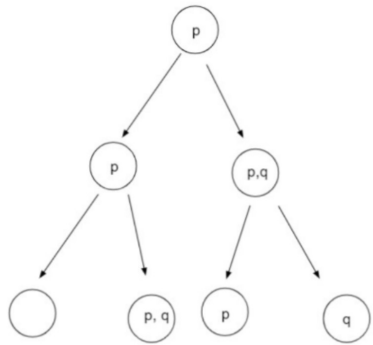

Figure 1: Kripke structure with example states and corresponding propositions. Each state is labeled with propositions *p* or *q*.

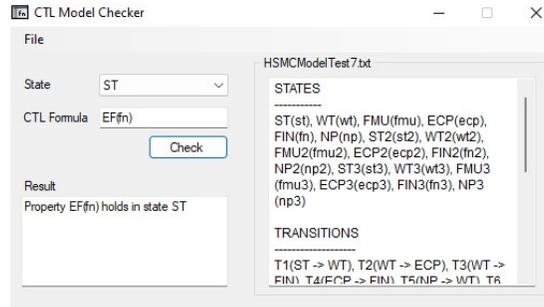

Figure 2: CTL Model Checker interface.

Figure 2 depicts the CTL Model Checker interface, in which the textual representation of the model, written in CTL, is first loaded into the application. The application then takes the start state as input. From here, a specific property is inputted and the CTL Model Checker will return whether or not it holds true, thus verifying the property. The model checker performs a backward propagation state space search to determine which states can be marked with a given property.

## 3.2 Interleavings

One way to model the concurrent execution of a set of steps by several agents is by the use of interleavings. An interleaving represents a possible permutation of the execution of concurrent steps. Thus, an initial approach to model concurrent execution is to enumerate all possible interleavings of execution of several steps by agents in a Kripke structure. Each interleaving must use unique state names to avoid the creation of loops. Interleavings do correspond to permutations of the same kind of steps so, the same step names do appear in different interleavings. To resolve the problem with loops we give unique names to different occurrences of the same step by attaching a number after the state name. For example, states VI1 and VI2 both represent the Visualization step in different interleavings. The number of interleavings is determined by the number of unique ways to order the $n$ concurrently executed steps. All interleavings combined contain $n! \times n$ states and $n! \times n$ transitions, thus contributing to the factorial explosion of the number of states and transitions, which has a higher growth rate than exponential explosion(refer to figure 3). The formula to determine the number of interleavings is below

$$n! = n \times (n - 1) \times (n - 2) \cdots \times 1$$

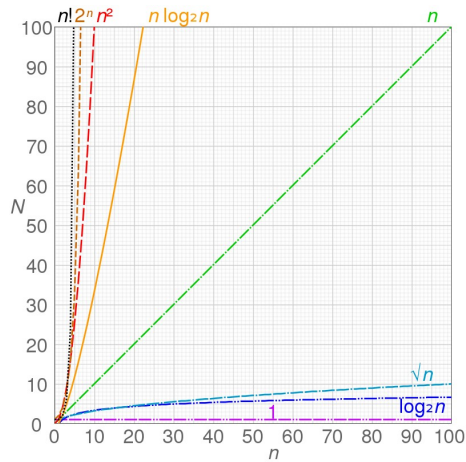

Figure 3: Commonly used formulas in Big O Notation

### 3.3   Big O notation(Upper Asymptotic Bound)

Big O notation is a function that serves as an upper bound to the running time function of the algorithm. The formula measures how the run time scales to the size of different inputs. To calculate the formula we first, go through our code line by line and determine whether or not that line execution time is constant no matter what our input is then, we assign these functions as $O(1)$. Then we look for functions in the code that increase linearly with the input and assign them $O(n)$. Finally, we look for functions that grow exponentially as the function increases and assign them $O(n^2)$. The next step is to drop all the $O(1)$ because we will not take them into consideration since their run time is not affected when we scale up the input. Lastly, we drop the non-dominant terms. For example, if we have $O(n + n^2)$ we drop the non-dominant term and change the formula to $O(n^2)$. In the case of model checking algorithms, we can use Big O notation to measure the run time as we increase the size of a CTL formula (amount of states, transitions, and sub-formulas).

## 4   Motivational Example

The use of Distributed Artificial Intelligence (AI) in Bioinformatics has grown more prevalent and it is necessary to be able to verify that the agents are communicating correctly especially in a field with no room for error. A common architecture for the implementation of a distributed AI system is a multi-agent system architecture. For instance, the process of genome annotation can benefit from the advantages of a multi-agent architecture. In figure 4, which gives the original description of the process from [**1**], the steps for different kinds of annotation are performed concurrently, so they can be executed by different agents. The concurrent execution of annotation steps for visualization, structural, functional, and community annotation can be modeled in a Kripke structure by enumerating their possible interleavings. The number of interleavings will be $n!$ where n is the number of concurrent agents, assuming each agent executes one step. Even though the diagram in figure 4 shows only one step for each kind of annotation it is most likely each one of those steps has an internal process whose detail is not shown in the figure. The Kripke structure that corresponds to the four annotation steps running concurrently is shown in figure 5. That Kripke structure can be used for analysis of the concurrent execution of the annotation steps.

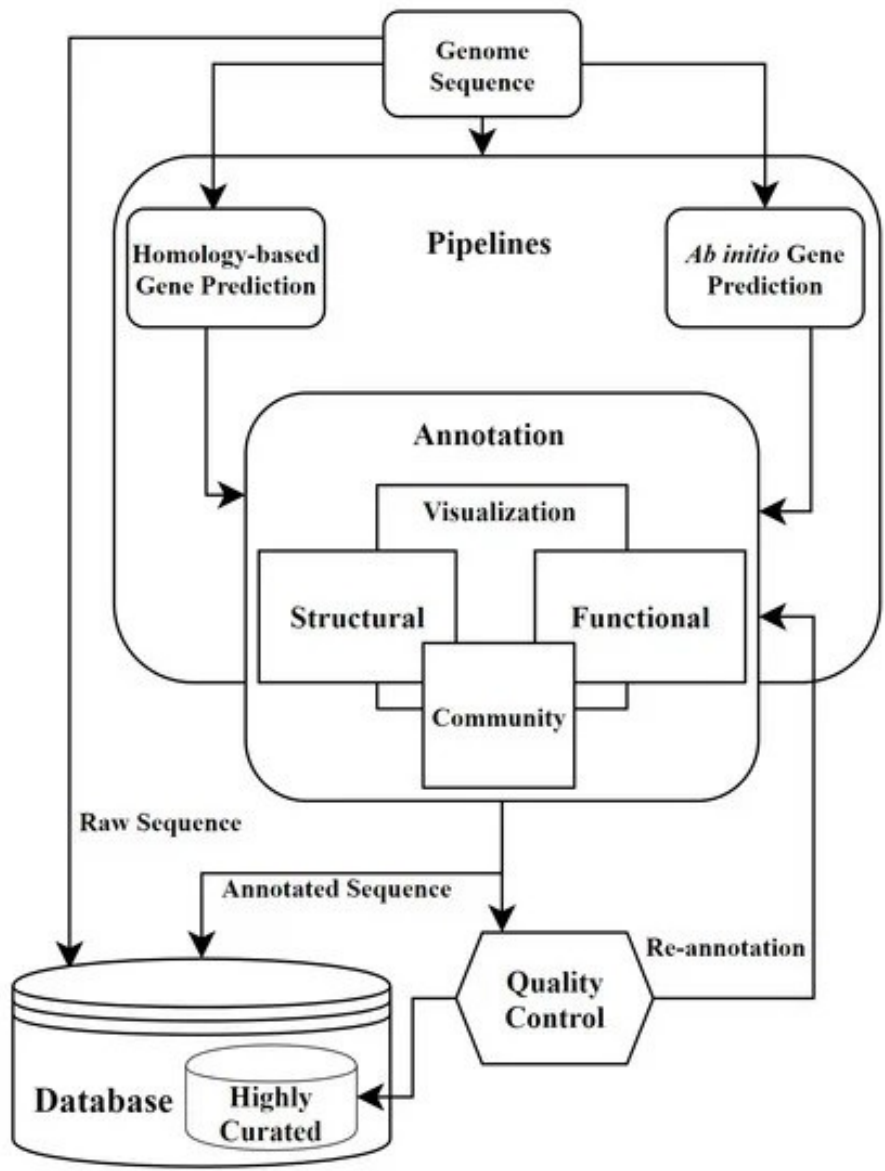

Figure 4: Workflow Diagram of Genome Annotation Process

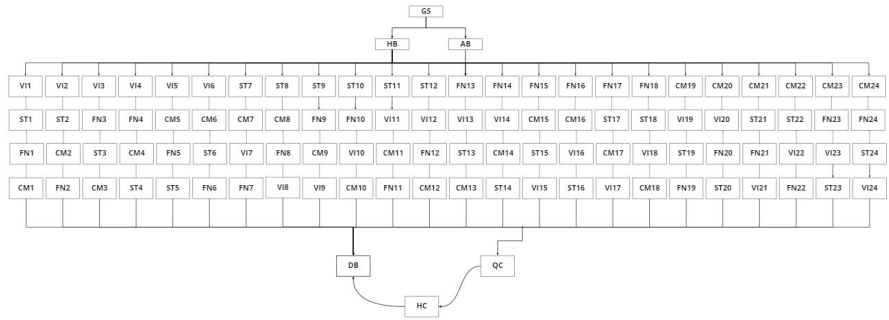

Figure 5: Kripke Structure of Workflow Diagram

# 5 Experimentation

## 5.1 Genome Annotation Process

Using the diagram of the workflow of genome annotation (shown in figure 4), it is possible to create a Kripke structure using the following states and transitions. The diagram begins with a genome sequence which can either go straight to the database or it can go to one of the prediction methods (Homology-based gene prediction and Ab initio Gene Prediction). Once the user chooses which method of prediction the genome sequence goes to the annotation phase where it goes through the following steps in any order: Visualizations, Community, Functional, and Structural. The diagram represents the 24 different permutations of the four steps. After annotations, the genome sequence can go to quality control which checks if the annotation was correct. If the annotation is correct then the genome sequence goes to the highly curated database. If the annotation is faulty then it goes back to annotation until it is correct. The genome sequence can also skip quality control and go straight to the database.

## 5.2 Errors in Genome Annotation Process

The process for genome annotation involves some steps concurrently executed by different agents. The processes for different agents are most likely written by different people, but the combined process must correspond to some common properties, so it is crucial that we can verify that the combined process does correspond to those common properties. In particular, the common property we are trying to verify in many of these cases is that the prediction states are done before annotation states. We must make sure that this holds regardless of the interleaving, that is, regardless of the order of execution of concurrent steps of a process.

### 5.2.1    Visualization Before Prediction Error

Figure 6 is the diagram of the faulty process where one of the annotation steps for Visualization, VI, is done before the prediction step AB. In Figure 6 we show the interleavings of three concurrently executed steps because the Visualization step is performed sequentially before step AB. The total number of interleavings is 6 and the number of steps is 18 and the number of transitions is 18. Notice that step VI is absent from the interleavings because it is not run concurrently when it is put before step AB. The interleavings are only done with three states which drastically reduces the number of interleavings to only 6 unique permutations. These interleavings exist to represent a group of steps running concurrently in each unique permutation. We formalized figure 6 into a textual representation of the Kripke structure and then gave it as input into a CTL model checker. Next, We ran the CTL formula AG(gs → AF(ab or hb) and ((ab or hb) → AF(vi1))). The property requires that if we reach step VI then we should have gone through step AB or HB. This property does not hold true in the initial step of the diagram, GS. This is the expected result because the Kripke structure in figure 6 has an error. This error could occur in a combined genome annotation process in which some steps are executed by different agents. It is very likely that the processes that correspond to those steps are written by different people, so it is possible that an error of this kind is introduced inadvertently. We used a CTL model checker to catch this error. Thus, we show that a process combined from different authors should be verified against common properties.

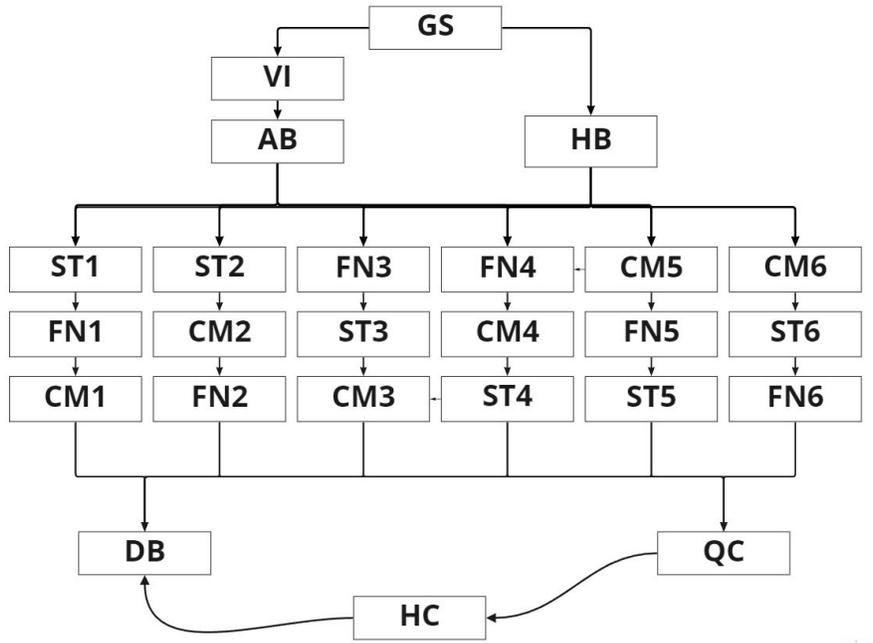

Figure 6: Kripke Structure of Visualization Before Prediction

## 5.3 Prediction After Annotation Error

Next, we show a faulty process in which one of the prediction steps was wrongly executed after the annotation steps. Figure 7 represents the case when one of the prediction states AB is done after all the annotation steps. In this figure, we show the concurrent execution of 4 steps which contains 24 interleavings, 96 steps, and 96 transitions in all the interleavings. Next, we submit the textual representation of this Kripke structure into a model checker and then verify it against the property below. In that property, we had to repeat the same pattern for all the representations of the Visualization step in each interleaving.

$AG(gs\rightarrow$ AF(ab or hb) and ((ab or hb) $\rightarrow$ AF(vi1 or vi2 or vi3 or vi4 or vi5 or vi6 or vi7 or vi8 or vi9 or vi10 or vi11 or vi12 or vi13 or vi14 or vi15 or vi16 or vi17 or vi18 or vi19 or vi20 or vi21 or vi22 or vi23 or vi24)))

This CTL formula represents this case corresponds to the property that any visualization step should be preceded by a prediction step. Step AB should have been executed before Visualization of every instance of the Visualization step in the interleavings. The CTL formula repeats the visualization preposition because it represents a condition that step AB or HB must be performed before each instance of the Visualization step for each interleaving. This repetition significantly increases the length of the CTL formula which is likely to increase the verification time. The expected result is that the property should not hold because the diagram shows that step AB is executed after all annotation steps. The CTL model checker shows that the property does not hold in step GS. The experimental results match our expectations. Thus, CTL model checking enables us to verify the model of concurrent execution of a group of steps in a process. Even though modeling of concurrency exponentially increases both the model and property sizes.

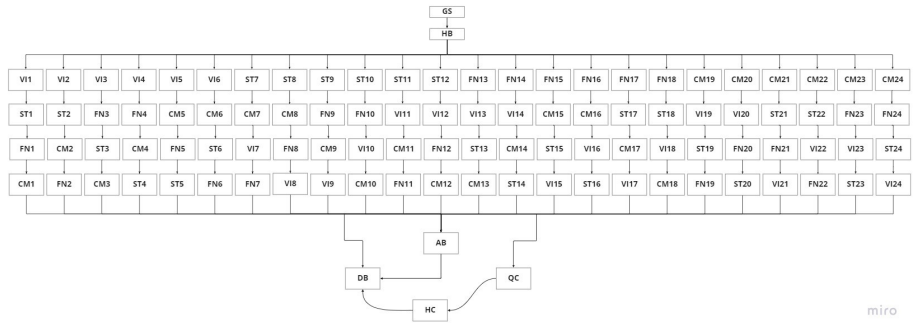

Figure 7: Kripke Structure of interleavings with error

# 6 Analysis

## 6.1 Results

This section presents the analysis of the experimental results. In this experiment, we applied the model checker to increasingly more structurally complex processes in which we varied the number of concurrently executed steps. The greater the number of interleavings according to the factorial or exponential dependence function of the running time on input size. The experimental results conform to the analytically expected dependence function.

## 6.2    Dependence of Running Time on Size of CTL Formula

The results shown in Figure 9 show a linear relationship between the length of the sub-formula and the verification time. This matches our prediction because the upper asymptotic bound is linear. It can be derived that the CTL model checker can run in polynomial time with any length of CTL formula, meaning Model checking can be performed in a reasonable time for any subformula length.

| t= ticks, 1 tick = 100 nanosec | | | | |
|---|---|---|---|---|
| CTL Formula Length (Subformulas) | Trial 1 time(t) | Trial 2 time(t) | Trial 3 time(t) | Average time(t) |
| 1 | 1250745 | 1367615 | 1216965 | 1250745 |
| 5 | 5190744 | 5171993 | 5153481 | 5190744 |
| 10 | 7647354 | 7809055 | 7546881 | 7647354 |
| 15 | 10244593 | 10198148 | 10162704 | 10244593 |
| 20 | 12560813 | 12488388 | 12549166 | 12560813 |

Figure 8: Influence of CTL Length on CTL Model Checker Time in Ticks

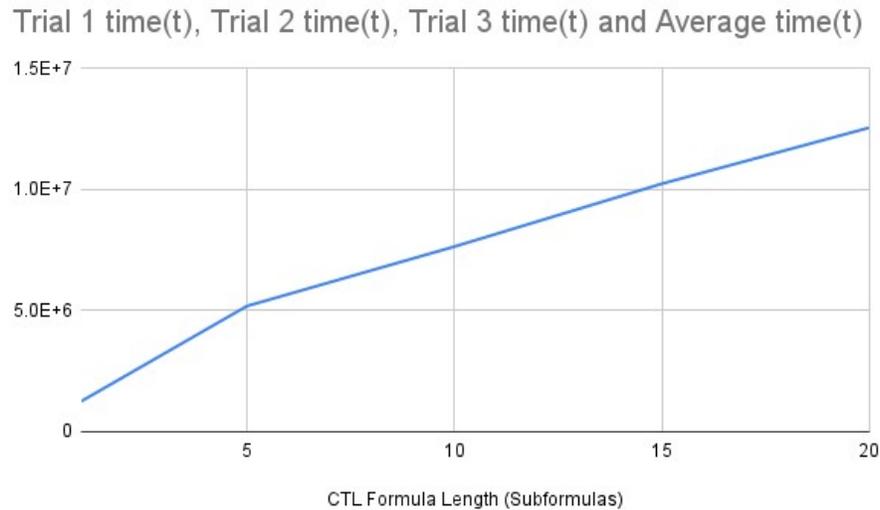

Figure 9: Graph of Time Dependence on CTL Formula Length

## 6.3   Dependence of Running Time on Size of Kripke Structure.

After running the CTL model checker to verify the genome annotation process, it was found that a variety of factors could be compared. Beginning with property types, the CTL approach analyzes and focuses on a sequence of events. For instance, the modification of the functional decomposition is quite apparent. In regards to computational complexity, the algorithms for the CTL model checking are linear with the size of the Kripke structure and the size of the input grows at the rate of *n*! where n is the number of agents executing one step each. The rate of growth of *n*! is worse(i.e. faster) than $2^n$. It is worse than an exponential explosion (Figure 11). Moreover, even though the computational complexity is linear, the occurrence of exponential explosion in the number of agents is inevitable. The main reason for its prevalence is due to the input size growing exponentially. Thus, this verification technique can only be applied to relatively small Kripke structures. The experimental results show a correlation between the average verification time (ms) and the size of the Kripke structure. In Figure 10 we see the data of the three Kripke structures of different sizes. The shape of the dependence graph is that of an exponential/factorial function. The graph shows the dependence of the running time of the model checker on the size of the input Kripke structure. The sizes of the Kripke structures that model concurrency via interleavings are significantly larger than those for the sequential execution of steps.

| t= ticks, 100 ticks = 1 nanosecond (C# tick) | | | | | | | | |
|---|---|---|---|---|---|---|---|---|
| Name of Kripke Structure | Transitions + Nodes | Trial 1 time(t) | Result | Trial 2 time(t) | Result | Trial 3 time(t) | Result | Average time (t) |
| One Interleaving Diagram | 21 | 98421 | TRUE | 97080 | TRUE | 94525 | TRUE | 96675 |
| Two Interleaving Diagram | 33 | 104429 | TRUE | 120524 | TRUE | 100378 | TRUE | 108444 |
| Annotation Before Prediction (Figure 6) | 67 | 198623 | FALSE | 196532 | FALSE | 198626 | FALSE | 197927 |
| Work Flow Diagram (Figure 5) | 274 | 1933180 | TRUE | 1959974 | TRUE | 1909480 | TRUE | 1934211 |
| Prediction After Annotation (Figure 7) | 274 | 3003642 | FALSE | 3055617 | FALSE | 3034083 | FALSE | 3031114 |

Figure 10: Time and Results from Kripke Structures

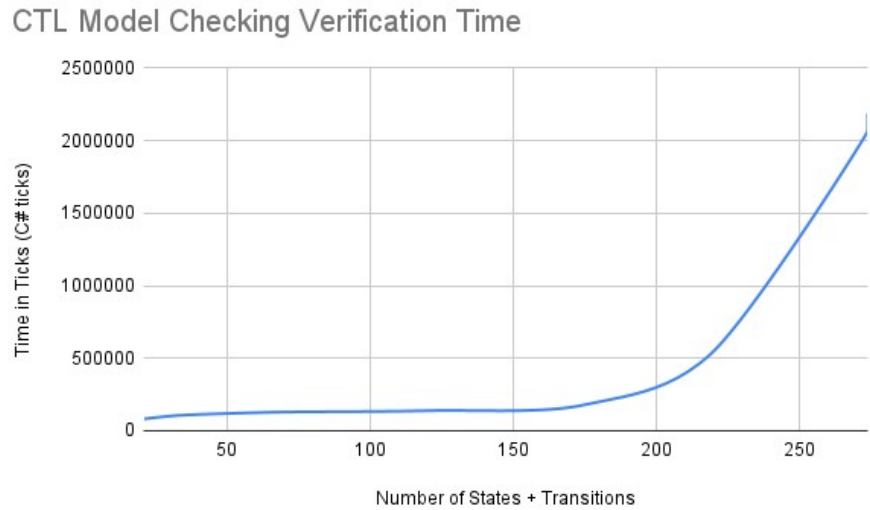

Figure 11: Graph of Time Dependence on Size of Kripke Structure

## 6.4    Upper Asymptotic Bound of Verification Time

The upper asymptotic bound (*BigO*) of CTL model checking algorithms is $O(|f| \times (|s|+|r|))$ where $|f|$ represents the number of sub-formulas in the property, $|s|$ represents the number of steps in a Kripke structure and $|r|$ represents the number of transitions in a Kripke structure. The linear time is attributed to the fact that CTL model checking algorithms are versions of state space search. For the Kripke structure in figure 6 and the property notation is $O(2 \times (31+42))$ and for figure 4 the Big O notation is $O(2 \times (102 + 172))$. Our experimental results do conform to these bounds.

# 7    Related Work

There are many papers that describe the application of Model Checking to combined Bioinformatics processes implemented as concurrent multi-agent systems. Thus, for related work, we searched for papers about the application of a multi agent approach to Bioinformatics processes. Most often these papers were written by experts in the area of Bioinformatics. In such papers, the authors would describe modeled Bioinformatics process. Multi-agent architecture of their system and experimental results related to the field of Bioinformatics. Not a single one of those papers attempted to verify their multi-agent systems against the correctness properties of the modeled process. Nevertheless we reviewed these papers as being close to the topic of Bioinformatics.

The paper by Girum Fitihamlak Ejigu and Jaehee Jung ([**1**]) gives a detailed explanation of the workflow of the genome annotation process and describes how the different agents should communicate with each other. The authors did not mention how they verified or tested their system. We used this paper to model the workflow diagram of the genome annotation process into a Kripke structure, then we turned the Kripke structure into a textual representation which we were able to use to validate properties against. After a literary search, we were not able to find any papers that validate the multi-agent genome annotation process using CTL model checking.

However, we were able to find a paper by Bert Bogaerts et al. ([**2**]) which focuses on using test data to validate the process of genome annotation. The paper provides a workflow diagram of the Whole Genome Sequencing process. Then, they used test data from the N. meningitidis reference strains to test if the process is annotating the expected results. They were able to validate that a specific process produces these expected results. The paper doesn't mention CTL model checking and lacks the advantage outlined in this paper.

In the paper by Tolga Ovatman et al. ([**3**]), the authors apply CTL model checking to Programmable logic controllers (PLCs). PLCs are a special type of computer that is capable of processing a large number of input and output operations within certain time constraints. The process is broken into 3 cycles: input data are read into memory, data in the memory are processed, and the output data are written. PLCs are necessary for real-time automation and have been applied to railway interlocking systems, nuclear power plants, and manufacturing conveyors. Errors in PLCs can be very dangerous and difficult to debug. Therefore it is necessary to employ verification techniques. The paper details the use of Model checking to verify PLCs and introduces the use of Petri nets while modeling processes.

In the paper by Francesco Maria Donini et al([**4**]), The authors apply CTL model checking to web applications. When creating a web application it is necessary to have multiple interactions between the database and the user. The authors were able to utilize Kripke structures to illustrate the processes necessary to create simple website features. For example, one step may be gathering login information and another step will be checking the login information in a database. The authors introduced errors into the process and then used CTL properties to catch the errors and ensure the steps are communicating correctly. They concluded that CTL model checking is an effective verification method for web applications

# 8 Conclusion

It is important to verify combined processes against common properties because most likely they will contain errors. These errors are most likely due to the fact that different parts of a combined process are written by different authors responsible for their domain of expertise. Multi-agent systems architecture is useful for the execution of bioinformatics processes. For instance, processes can be executed faster due to internal concurrency. Testing for faults in processes with internal concurrency delivers a low level of assurance because testing is always sampling and because testing does not enforce the enumeration of all possible interleavings. One of the great advantages of static analysis methods, such as model checking, is that by enumeration of all possible interleavings, they greatly increase their assurance of analysis results. One of the disadvantages that result from this enumeration is the exponential explosion of the size of the model. Our experimental results confirm both the high level of assurance of the results of the analysis a concurrent process by model checking and the exponential time of the analysis. This system for the execution of Bioinformatics processes that have internal concurrency can greatly benefit from verification by model checking or other static analysis methods. In the reviewed related literature about Bioinformatics processes we have not come across a single instance of verification by a static analysis method which is most likely due to the fact that experts in the field of Bioinformatics are not aware of such methods.

Multidisciplinary research in the field of verification of Bioinformatics processes can be beneficial to both the field of Bioinformatics and the field of mathematically rigorous program analysis. The main contribution of our work is the application of combined Bioinformatics processes. The Kripke structure for the combined processes is created automatically. The experimental results show that there is an exponential explosion in the size of the Kripke structure due to the number of concurrent processes. There are a number of ways to mitigate this exponential explosion. For instance, the use of a more compact representation, such as the use of binary representation, on-the-fly model checking, distribution using concurrent versions of model checking algorithms, reducing Kripke structure size by further abstraction. In the future, we would like to investigate these techniques.

The impact of our approach that addresses these existing limitations is that we use a theoretical workflow diagram to identify errors that may be created when the individually created processes are combined by one author. Rather than having to assemble the process to verify it, our approach allows for verification to be done before assembly.